\documentclass[10pt,conference,letterpaper]{IEEEtran}

\usepackage{times,amsmath}
\usepackage{amsfonts}
\usepackage{amssymb,bm}
\usepackage{bbm}
\usepackage{algorithm}
\usepackage{algpseudocode}
\usepackage{cite}
\usepackage{graphicx}
\usepackage{epstopdf}
\usepackage[caption=false,font=footnotesize]{subfig}
\usepackage{fixltx2e}
\usepackage{balance}
\usepackage{cases}
\usepackage{color}

\allowdisplaybreaks

\usepackage{amsthm}
\newtheorem{theorem}{Theorem}
\newtheorem{definition}{Definition}
\newtheorem{lemma}{Lemma}
\newtheorem{remark}{Remark}

\newtheorem{corollary}{Corollary}
\usepackage{chngcntr}
\counterwithout{theorem}{section}
\counterwithout{definition}{section}
\counterwithout{lemma}{section}
\counterwithout{remark}{section}
\counterwithout{assumption}{section}
\counterwithout{proposition}{section}
\counterwithout{corollary}{section}

\DeclareMathOperator{\E}{\mathbbmss{E}}
\DeclareMathOperator{\Prob}{\mathbbmss{P}}

\IEEEoverridecommandlockouts
\begin{document}
\bstctlcite{IEEEexample:BSTcontrol}
%
\title{On the DoF of Parallel MISO BCs with Partial CSIT: Total Order and Separability}

\author{
\IEEEauthorblockN{Hamdi Joudeh and Bruno Clerckx}
\fontsize{9}{9}\upshape
Department of Electrical and Electronic Engineering, Imperial College London, United Kingdom \\
\fontsize{9}{9}\selectfont\ttfamily\upshape
Email: \{hamdi.joudeh10, b.clerckx\}@imperial.ac.uk
\thanks{This work is partially supported by the U.K. Engineering and Physical Sciences Research Council (EPSRC) under grant EP/N015312/1.}
}
\maketitle

\begin{abstract}
We study the degrees of freedom (DoF) of a $K$-user parallel MISO broadcast channel with arbitrary levels of partial CSIT over each subchannel.
We derive a sum-DoF upperbound which depends on the average CSIT quality of each user.
This upperbound is shown to be tight under total order, i.e. when the order of users with respect to their CSIT qualities is preserved over all subchannels.
In this case, it is shown that separate coding over each subchannel is optimum in a sum-DoF sense.
\end{abstract}


\section{Introduction}
\newcounter{Theorem_Counter}
\newcounter{Proposition_Counter} 
\newcounter{Lemma_Counter} 
\newcounter{Remark_Counter} 
\newcounter{Assumption_Counter}
\newcounter{Definition_Counter}
Broadcast channels (BCs) model the downlink of wireless systems consisting of one transmitter and multiple uncoordinated receivers.
One class of BCs that received considerable attention over the last one and a half decades is the MIMO BC. The theoretical breakthroughs in characterizing the capacity region for a subclass of MIMO BCs, particularly those with private messages and perfect Channel State Information at the Transmitter (CSIT), continue to have significant practical impact today. Indeed, downlink multiuser (MU) MIMO -- the practical counterpart of the MIMO BC -- plays an integral role in current and future wireless systems and standards \cite{Clerckx2013}.

One of the important issues that arise in practice is the inability to provide highly accurate and up-to-date CSIT. While it is understood that the MIMO BC is very sensitive to such inaccuracies, fundamental limits under such conditions remain largely a mystery.
Such problems proved to be highly elusive even in the most basic settings and under simplifying assumptions and approximations.
For example, a conjecture on the collapse of the Degrees of Freedom (DoF) in the $2$-user MISO BC -- where each user is equipped with a single antenna -- under finite precision CSIT remained open for almost a decade \cite{Lapidoth2005}. This was only recently settled by Davoodi and Jafar in \cite{Davoodi2016}. The key ingredient of the proof is the Aligned Image Sets (AIS) approach, a combinatorial argument that bounds the relative size of images cast by a set of codewords at different receivers under CSIT uncertainty.
The AIS approach proved to be powerful in a more general sense, providing a tight sum-DoF upperbound for the $K$-user MISO BC with arbitrary levels of partial CSIT, where the CSIT error for the $k$-th user scales as $\mathrm{SNR}^{-\alpha_{k}}$ for some quality parameter $\alpha_{k} \in [0,1]$.
Achievability of the upperbound was shown using signal space partitioning and rate-splitting \cite{Davoodi2016a,Joudeh2016}.

\emph{The parallel MISO BC:} This is a generalization in which transmission is carried out over orthogonal subchannels, where each subchannel can be viewed as a MISO BC in its own right.
Parallel channel models are motivated by frequency-selective or time-varying wireless channels, in which each subchannel represents an orthogonal frequency tone in the former or a collection of time instances with constant channel in the latter.
An important issue that arises when studying parallel channels is separability.
In general, this refers to the optimality of separate coding across different subchannels subject to a joint power constraint \cite{Cadambe2009}.
Parallel MISO BCs are known to be separable when perfect CSIT is available \cite{Mohseni2006}.

Moving towards imperfect CSIT, the parallel MISO BC has been studied within the DoF framework.
In \cite{Tandon2013}, the alternating CSIT setting was introduced, in which the transmitter acquires
perfect ($\mathrm{P}$), delayed ($\mathrm{D}$) or no ($\mathrm{N}$) knowledge of each user's channel,
and such knowledge varies over time instances according to some probabilities.
As a result, the joint CSIT of all users alternates between a finite number of states, and each of such states can be considered as a subchannel (or a collection of subchannels) in a parallel MISO BC.
The results in \cite{Tandon2013} were primarily focused on $2$-user networks.
The alternating setting was further investigated for $K$-user networks in \cite{Rassouli2016}.
The $2$-user parallel MISO BC with partial CSIT was studied in \cite{Hao2013}. Here the transmitter acquires an arbitrary level of partial instantaneous knowledge of each user's channel over each subchannel, and  partial CSIT follows the model in \cite{Davoodi2016}.
In this paper, we consider the $K$-user parallel MISO BC with arbitrary partial CSIT levels.
This setting is seemingly most related to the one considered in \cite{Hao2013}.
However, close, yet perhaps less obvious, ties with the alternating setting in \cite{Rassouli2016} also exist as highlighted in Section \ref{sec:main results}.
One important conclusion that can be made from the aforementioned works above is that parallel MISO BCs with partial CSIT are inseparable in general.
This is best exemplified by the $2$-user/$2$-subchannel $\mathrm{PN}$-$\mathrm{NP}$ setting,
in which perfect CSIT is available for one user only in the first subchannel, and for the other user only in the second subchannel.
It was shown in \cite{Tandon2013} that joint coding achieves the optimum sum-DoF in this setting, which is strictly greater than the sum-DoF achieved through separate coding.
Note that $\mathrm{P}$ and $\mathrm{N}$ are special cases of partial CSIT.
Other examples in more sophisticated settings can be found in \cite{Rassouli2016,Hao2013,Chen2013}.

\emph{Total Order and Separability:}
Separability of parallel channels is a desirable feature, as it simplifies coding and multiple access in multiuser setups.
Hence, a question that arises is whether MISO BCs with partial CSIT are separable under certain conditions.
We answer this with the affirmative by showing that parallel MISO BCs with partial CSIT are
indeed separable in the sum-DoF sense when users abide a total order condition.
Total order holds if and only if CSIT qualities for users have the same order across all subchannels.
In other words, for any pair of users $i,j$ and subchannels $m,l$, having $\alpha_{i}^{[m]} \geq \alpha_{j}^{[m]}$ implies
$\alpha_{i}^{[l]} \geq \alpha_{j}^{[l]}$.
To show the separability result, we derive an upperbound for the sum-DoF of the MISO BC with arbitrary levels of partial CSIT.
The upperbound is expressed in terms of the average CSIT qualities across subchannels, and does not assume or require any order of users.
However, order is somehow embedded through functional dependencies imposed as part of the AIS approach.
We show that this upperbound is tight under total order, and that it is achieved through separate coding over subchannels.
Since the upperbound is general, the result implies that given the average CSIT qualities, a total order of users achieves the highest possible sum-DoF.
Some insights are drawn from the sum-DoF result including $\mathrm{PN}$ decomposability, where we show that the setup at hand is equivalent to a parallel MISO BC with CSIT qualities drawn from $\{\mathrm{P},\mathrm{N}\}$.
Moreover, we discuss various possible extensions to the work including the characterization of the DoF region and tackling the general non-ordered case.

\emph{Notation:}
The following notations are used in the paper.
$a,A$ are scalars ($A$ usually denoting a random variable), $\mathbf{a}$ is a column vector (unless specifically stated otherwise) and $\mathbf{A}$ is a matrix.
$(a_{1},\ldots,a_{K})$ denotes a $K$-tuple of scalars.
For any positive integer $K$, the set $\{1,\ldots,K\}$ is denoted by $\langle K \rangle$.
\section{System Model}
\label{sec:system model}
Consider a MISO BC in which a transmitter with $K$ antennas communicates independent messages to $K$ single-antenna users (receivers) indexed by the set $\langle K \rangle$.
The channel has $M$ parallel subchannels indexed by $\langle M \rangle$.
For a transmission taking place over $n>0$ channel uses, the signal received by the $k$-th user through the $m$-th subchannel at the $t$-th channel use, where $t \in \langle n \rangle$, is given by
\begin{equation}
\label{eq:received signal}
y_{k}^{[m]}(t)= \mathbf{h}_{k}^{[m]}(t) \mathbf{x}^{[m]}(t)+ z_k^{[m]}(t)
\end{equation}
in which $\mathbf{h}_{k}^{[m]}(t) \in \mathbb{C}^{1 \times K}$ is the fading channel vector,
$\mathbf{x}^{[m]}(t) \in \mathbb{C}^{K\times 1}$ is the vector of transmitted symbols and $z_k^{[m]}(t) \sim \mathcal{N}_{\mathbb{C}}(0,1)$ is the additive white Gaussian noise (AWGN).
Note that users are indexed by the subscript, subchannels are indexed by the superscript in square brackets and channel uses are indexed by the argument in the round brackets.
In a slight abuse of notation, $\mathbf{h}_{k}^{[m]}(t)$ is used to denote a row vector to avoid the  otherwise cumbersome notation.
To avoid degenerate situations, the ranges of values of all channel coefficients are bounded away from zero and infinity which is not
a major restriction as explained in \cite{Davoodi2016}.
The transmitted signal is subject to the power constraint
\begin{equation}
\frac{1}{nM} \sum_{t=1}^{n} \sum_{m=1}^{M} \| \mathbf{x}^{[m]}(t) \|^{2} \leq P
\end{equation}
which can be interpreted as the average power per-channel-use per-subchannel.
\subsection{Partial CSIT}
\label{subsec:partial CSIT}
Under partial CSIT, the channel vectors $\mathbf{h}_{k}^{[m]}(t)$, for all $k \in \langle K \rangle$ and $m \in \langle M \rangle$, are modeled as
\begin{equation}
\label{eq:channel model}
\mathbf{h}_{k}^{[m]}(t) = \hat{\mathbf{h}}_{k}^{[m]}(t) + \sqrt{P^{-\alpha_{k}^{[m]}}}\tilde{\mathbf{h}}_{k}^{[m]}(t)
\end{equation}
where $\hat{\mathbf{h}}_{k}^{[m]}(t) \in \mathbb{C}^{1 \times K}$ is the channel estimate and
$\tilde{\mathbf{h}}_{k}^{[m]}(t) \in \mathbb{C}^{1 \times K}$ is the estimation error.
We assume a non-degenerate channel uncertainty model where the entries of $\hat{\mathbf{h}}_{k}^{[m]}(t)$ and $\tilde{\mathbf{h}}_{k}^{[m]}(t)$ are drawn from continuous joint distributions with bounded densities, with the difference that the realizations of
$\hat{\mathbf{h}}_{k}^{[m]}(t)$  are revealed to the transmitter while the realizations of $\tilde{\mathbf{h}}_{k}^{[m]}(t)$ are not available to the transmitter \cite{Davoodi2016a}.
This partial CSIT is parameterized by $\alpha_{k}^{[m]}$,
which measures the quality of the estimate available to the transmitter for the $k$-th user over the $m$-th subchannel.
It is assumed that $\alpha_{k}^{[m]} \in [0,1]$ for all $k \in \langle K \rangle$ and $m \in \langle M \rangle$, which measures the whole range of partial CSIT in a DoF sense.
$\alpha_{k}^{[m]} = 0$ corresponds to the case where current CSIT is not known and $\alpha_{k}^{[m]} = 1$ corresponds to perfectly known CSIT, both in a DoF sense\footnote{Note that the unknown CSIT case also includes finite-precision CSIT, which is shown to be unuseful in a DoF sense \cite{Davoodi2016}.}.
Such extreme states are denoted by $\mathrm{N}$ and $\mathrm{P}$  respectively.
CSIT qualities for the $K$ users over the $m$-th subchannel are given in the tuple
$\bm{\alpha}^{[m]} = ( \alpha_{1}^{[m]} ,\ldots,\alpha_{K}^{[m]})$.
On the other hand, the $M$ CSIT qualities associated with user-$k$ over all subchannels are given in the tuple $\bm{\alpha}_{k} = (\alpha_{k}^{[1]} ,\ldots,\alpha_{k}^{[M]} )$. The corresponding average quality is given by $\alpha_{k} = \frac{1}{M} \sum_{m=1}^{M} \alpha_{k}^{[m]}$.
Average CSIT qualities for all $K$ users are given by the tuple $\bm{\alpha} = (\alpha_{1} ,\ldots,\alpha_{K} )$. Without loss of generality, we assume that
 \begin{equation}
\label{eq:average CSIT order}
\alpha_{1} \geq \alpha_{2} \geq \cdots \geq \alpha_{K}.
\end{equation}
\subsection{Messages, Rates, Capacity and DoF}
\label{subsec:msgs-rates-capacity-DoF}
The transmitter has messages $W_1,\ldots, W_K$ intended to the corresponding users.
Codebooks, probability of error, achievable rate tuples $(R_1(P),\ldots,R_K(P))$ and the capacity region $\mathcal{C}(P)$ are all defined in the standard Shannon theoretic sense.
Note that achievable rates are defined as $n \rightarrow \infty$, yet $M$ remains fixed for a given setup.
The DoF tuple $(d_1, \ldots, d_K)$ is said to be achievable if there exists $(R_1(P),\ldots,R_K(P)) \in \mathcal{C}(P)$ such that $d_k=\lim_{P \to \infty} \frac{R_k(P)}{M\log(P)}$ for all $k \in \langle K \rangle$.
Here $M\log(P)$ approximates the baseline capacity of $M$ subchannels at high SNR.
The DoF region is defined as the closure of all achievable DoF tuples $(d_1, \dots, d_K)$, and is denoted by $\mathcal{D}$.
The optimum sum-DoF (or simply DoF) value is defined as $d_{\Sigma} = \max_{(d_1, \dots, d_K) \in \mathcal{D}} \sum_{i=1}^{K} d_{i}$.
\begin{remark}
\label{remark:DoF per subchannel}
According to the above definition, the considered DoF is per-channel-use per-subchannel.
For example, if channel uses and subchannels represent time instances and orthogonal frequency tones respectively,
the DoF represents the number of interference free spatial dimensions per orthogonal time-frequency signalling dimension at high SNR.
\end{remark}
\section{Main Results and Insights}
\label{sec:main results}
We start by defining a total order of users.
\begin{definition}
\label{def:total order}
Users are totally ordered with respect to their CSIT qualities if there exists a permutation $\pi(\cdot)$ of size $K$ over the set $\langle K \rangle$ such that $\bm{\alpha}_{\pi(1)} \geq \bm{\alpha}_{\pi(2)} \geq \cdots \geq \bm{\alpha}_{\pi(K)}$, where the vector inequalities are element-wise.
For the average CSIT order in \eqref{eq:average CSIT order}, the total CSIT order condition becomes
\begin{equation}
\label{eq:total CSIT order}
\bm{\alpha}_{1} \geq \bm{\alpha}_{2} \geq \cdots \geq \bm{\alpha}_{K}.
\end{equation}
\end{definition}
Now we state the main result of this paper.
\begin{theorem}
\label{theorem:sum DoF}
For the parallel MISO BC described in Section \ref{sec:system model}, the sum-DoF is bounded above as
\begin{equation}
\label{eq:sum DoF}
d_{\Sigma} \leq 1 + \alpha_{2} + \cdots + \alpha_{K}.
\end{equation}
Furthermore, this bound is achievable when users are totally ordered as described in Definition \ref{def:total order}.
\end{theorem}
It is evident that the upperbound in \eqref{eq:sum DoF} depends only on the average CSIT qualities $\bm{\alpha}$.
The implication of this under total order is emphasised in Section \ref{subsec:PN decomposition}.
Next, we recover special cases of Theorem \ref{theorem:sum DoF} from previous works.
\subsection{Relation to Previous Results}
\subsubsection{A single subchannel}
\label{subsec:single subchannel}
In this case, CSIT qualities coincide with the average qualities $\alpha_{1},\ldots,\alpha_{K}$ and the upperbound in \eqref{eq:sum DoF} reduces to the one in \cite{Davoodi2016}.
Moreover, the upperbound is achieved using a rate-splitting strategy \cite{Joudeh2016}.
Note that in this case, a total order of users naturally exists.
\subsubsection{The two-users case}
The upperbound in \eqref{eq:sum DoF}  boils down to $d_{1} + d_{2}\leq 1 + \alpha_{2}$,
which depends only on the smaller average CSIT quality $\alpha_{2}$.
In this special case, total order of users is not necessary as
the upperbound can be achieved using an augmented\footnote{Augmentation is carried out over subchannels in a space-time/frequency manner, depending on the setup.} rate-splitting scheme \cite{Chen2013,Hao2013}.
\subsubsection{Alternating current CSIT}
\label{subsubsec:alternating CSIT}
This corresponds to the setup in \cite{Rassouli2016}
with zero probability for states with delayed CSIT.
We describe this briefly here to better see the relationship with the model described in Section \ref{sec:system model}.
The transmitter may have perfect ($\mathrm{P}$) or no ($\mathrm{N}$) knowledge of the CSI for each user and such knowledge varies over time instances (or channel uses $t$) according to some probabilities.
In particular, the joint CSIT state denoted by $\mathbf{s}$ takes one of $2^{K}$ possible values, each associated with a probability $\lambda^{\mathbf{s}}$ such that $\sum_{\mathbf{s}} \lambda^{\mathbf{s}} = 1$.
From the joint probabilities, the marginal probabilities of $\mathrm{P}$ and $\mathrm{N}$ for user-$k$ are given by $\lambda_{k}^{\mathrm{P}} = \sum_{\mathbf{s}:s_{k} = \mathrm{P}} \lambda^{\mathbf{s}}$ and $\lambda_{k}^{\mathrm{N}} = \sum_{\mathbf{s}:s_{k} = \mathrm{N}} \lambda^{\mathbf{s}} = 1- \lambda_{k}^{\mathrm{P}}$ respectively\footnote{Considering a $2$-user setup for example, the joint CSIT state is drawn from $\left\{(\mathrm{P,P}),(\mathrm{P,N}),(\mathrm{N,P}),(\mathrm{N,N}) \right\}$ with probabilities $\left\{\lambda^{(\mathrm{P,P})},\lambda^{(\mathrm{P,N})},\lambda^{(\mathrm{N,P})},\lambda^{(\mathrm{N,N})}\right\}$ respectively. The marginal probabilities of user-1 are $\lambda_{1}^{\mathrm{P}} = \lambda^{(\mathrm{P,P})} + \lambda^{(\mathrm{P,N})}$
and $\lambda_{1}^{\mathrm{N}} = \lambda^{(\mathrm{N,P})} + \lambda^{(\mathrm{N,N})}$.}.
This can be seen as a special case of the considered setup by treating each joint state $\mathbf{s}$ as a collection of subchannels\footnote{\label{footnote:alternating}While drawing such parallels, a causality issue arises for schemes with inter-subchannel dependencies.
This can be resolved by imposing an i.i.d evolving assumption and a block-Markov modification \cite[Remark 7]{Tandon2013}.}.
Specifically, we associate each $\mathbf{s}$ with $M^{\mathbf{s}}$ subchannels, indexed by $\mathcal{M}^{\mathbf{s}} \subseteq \langle M \rangle$, such that $\sum_{\mathbf{s}} M^{\mathbf{s}} = M$.
For each subchannel $m$, CSIT qualities are obtained by mapping $\mathbf{s}$ into $\bm{\alpha}^{[m]}$ with entries drawn from $\{0,1\}$, which remain fixed over all $m \in \mathcal{M}^{\mathbf{s}}$.
By making $M$ sufficiently large, any set of joint probabilities can be realized by taking $\lambda^{\mathrm{s}} = \frac{M^{\mathrm{s}}}{M}$.
This corresponds to the portion of subchannels in which the joint state is $\mathbf{s}$.
Moreover, the marginal probability $\lambda_{k}^{\mathrm{P}}$ is now equivalent to $\alpha_{k}$, which corresponds to the portion of subchannels in which $s_{k} = \mathrm{P}$ (or $\alpha_{k}^{[m]} =1$). It follows that \eqref{eq:sum DoF} reduces to the sum-DoF upperbound in \cite[Th. 1]{Rassouli2016}.
\subsection{Separability}
An important implication of Theorem \ref{theorem:sum DoF} is the sum-DoF separability.
This type of separability is defined as follows.
\begin{definition}
Sum-DoF separability holds if and only if
\begin{equation}
\label{eq:sum-DoF separability}
d_{\Sigma} = \frac{1}{M} \sum_{m=1}^{M} d_{\Sigma}^{[m]}
\end{equation}
where $d_{\Sigma}^{[m]}$ is the sum-DoF of the $m$-th subchannel when considered as a separate network.
\end{definition}
Under such separability, the sum-DoF is achieved by separate coding with power allocation $P$ for each subchannel.
\begin{corollary}
The subchannels of the parallel MISO BC described in Section \ref{sec:system model} with totally ordered partial CSIT according to Definition \ref{def:total order} are sum-DoF separable.
\end{corollary}
This is easily concluded from Theorem \ref{theorem:sum DoF} as follows. Under total order, equality holds in \eqref{eq:sum DoF}. This in turn is equal to the right-hand-side of \eqref{eq:sum-DoF separability}, as $d_{\Sigma}^{[m]} = 1 + \alpha^{[m]}_{2}+\cdots+\alpha^{[m]}_{K}$
for the $m$-th subchannel. This is further elaborated in Section \ref{subsec:single subchannel}.
\subsection{$\mathrm{PN}$ Decomposition}
\label{subsec:PN decomposition}
\begin{figure}[t!]
\centering
\includegraphics[width = 0.48\textwidth]{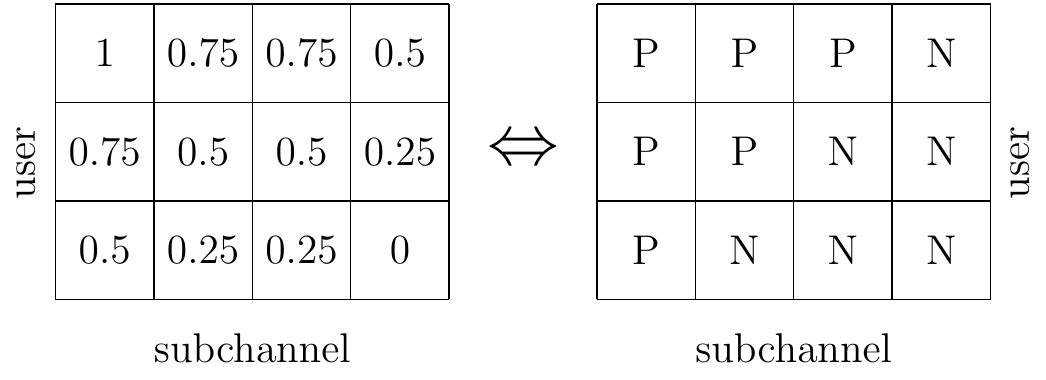}
\caption{Example of $3$-user/$4$-subchannel settings with partial CSIT (left) and $\mathrm{P}\mathrm{N}$ CSIT (right). The average CSIT qualities are the same for both settings, i.e. $\bm{\alpha} = \left(\frac{3}{4},\frac{1}{2},\frac{1}{4}\right)$.
Total order holds in both settings. Hence, the sum-DoF is the same and is given by $d_{\Sigma} = \frac{7}{4}$.}
\label{Fig_CSIT_pattern}
\end{figure}

Let $\mathrm{P}^{l}\mathrm{N}^{K-l}$ be the joint state in which the CSIT is perfectly known for the first $l$ users and not known for the remaining $K-l$ users.
For notational briefness, we simply denote this state by $\mathrm{P}^{l}$.
Mapping $\mathrm{P}^{l}$ to the corresponding tuple of CSIT qualities,
it is easily seen that the sum-DoF of a MISO BC with such state is given by $d_{\Sigma}^{\mathrm{P}^{l}} = \max\{1,l\}$.
Moreover, under total order, the sum-DoF in \eqref{eq:sum DoF} can be expressed by
\begin{equation}
\label{eq:sum DoF PN decomposition}
d_{\Sigma} = w_{0} d_{\Sigma}^{\mathrm{P}^{0}} + w_{1} d_{\Sigma}^{\mathrm{P}^{1}} + \dots + w_{K} d_{\Sigma}^{\mathrm{P}^{K}}
\end{equation}
where $w_{l} = \alpha_{l} - \alpha_{l+1}$ for all $l \in \{0,1,\ldots,K\}$, $\alpha_{0} = 1$ and $\alpha_{K+1} = 0$.
The decomposition in \eqref{eq:sum DoF PN decomposition} is inline with the weighted-sum interpretation in \cite{Hao2013,Hao2017} and the notion of signal-space partitioning in \cite{Davoodi2016a,Yuan2016}.
Under total order and in terms of the sum-DoF, the parallel MISO BC with partial CSIT is equivalent to a parallel MISO BC where
the admissible CSIT state for each user in any subchannel belongs to $\{\mathrm{P},\mathrm{N}\}$ (or equivalently, $\alpha_{k}^{[m]} \in \{1,0\}$ for all $k \in \langle K \rangle$ and $m \in \langle M \rangle$).
In the equivalent parallel channel, the weight $w_{l}$ is the fraction of subchannels over which the joint state is $\mathrm{P}^{l}$.
On the other hand, $\sum_{l = k}^{K} w_{l}$ is the fraction of subchannels over which perfect CSIT for user-$k$ is available.
By observing that $\sum_{l = k}^{K} w_{l} = \alpha_{k}$,
it can be seen that as long as total order is preserved,
reporting partial CSIT over all subchannels with average quality $\alpha_{k}$
is equivalent to reporting perfect CSIT over a fraction $\alpha_{k}$ of subchannels, and no CSIT over the remaining subchannels.
An illustrative example is shown in Fig. \ref{Fig_CSIT_pattern}.
This can have an operational significance for example in OFDMA systems where CSIT feedback is carried out over a subset of subchannels only.
\begin{remark}
We have previously stated that under mild conditions (see footnote \ref{footnote:alternating}),
any MISO BC with alternating current CSIT is equivalent to a parallel MISO BC with partial CSIT.
The $\mathrm{PN}$ decomposition shows that the reverse statement holds for the sum-DoF under total order.
In particular, the fractions $w_{l}$ and $\sum_{l = k}^{K} w_{l}$ can be interpreted as the joint probability $\lambda^{\mathrm{P}^{l}}$  and the marginal probability $\lambda_{k}^{\mathrm{P}}$, respectively.
Hence, the sum-DoF of the parallel MISO BC is is equivalent to the sum-DoF of a MISO BC with alternating current CSIT.
\end{remark}
\section{Proof of Theorem \ref{theorem:sum DoF}}
\label{sec:proof of theorem}
\subsection{Achievability}
Achievability is based on separate coding over each subchannel and involves a two-layer rate-splitting scheme.
The first layer divides each message into $M$ submessages corresponding to the $M$ subchannels, e.g.
$W_{k} \mapsto \left( W_{k}^{[1]},\ldots W_{k}^{[M]}\right)$.
This yields $M$ groups of $K$ submessages, each to be transmitted independently over a corresponding subchannel.
The second layer of rate-splitting is carried out independently for each subchannel.
Taking the $m$-th subchannel for example, the rate-splitting scheme in \cite{Joudeh2016} is used
where transmission powers of the common and private parts are adjusted according to the maximum CSIT quality
$\max \bm{\alpha}^{[m]}$.
This maximizes the sum-DoF achieved over this subchannel, given by\footnote{The normalization by $M$ is due to the fact that we are considering the DoF per subchannel as pointed out in Remark \ref{remark:DoF per subchannel}.}
$\sum_{i=1}^{K}d_{i}^{[m]} = \frac{1}{M}\left(1 + \sum_{i=1}^{K} \alpha_{i}^{[m]} - \max \bm{\alpha}^{[m]} \right)$.
The achievable sum-DoF over all subchannels is given by
\begin{equation}
\label{eq:achievable sum-DoF}
\sum_{i=1}^{K}\sum_{m=1}^{M} d_{i}^{[m]} =   1 + \sum_{i=1}^{K} \alpha_{i} - \frac{1}{M} \sum_{m=1}^{M} \max \bm{\alpha}^{[m]}.
\end{equation}
It can be easily checked that \eqref{eq:achievable sum-DoF} is equal to the right-hand-side of \eqref{eq:sum DoF} under total order.
\subsection{Converse}
The converse follows in the footsteps of \cite{Davoodi2016}, with slight differences due to the parallel channel which are highlighted along the way.
For simplicity, we focus on real channels. The extension to complex channels is cumbersome yet conceptually straightforward as shown in \cite{Davoodi2016}.
The first step is to reduce the number of channel parameters through a canonical transformation, performed in this case for each subchannel.
This is followed by discretization. The input-output relationship of the resulting deterministic channel is given by
\begin{equation}
\bar{Y}_{k}^{[m]}(t) = \bar{X}_{k}^{[m]}(t) + \sum_{i = 1}^{k-1} \bigl\lfloor G_{ki}^{[m]}(t) \bar{X}_{i}^{[m]}(t) \bigr\rfloor
\end{equation}
where  $\bar{X}_{k}^{[m]}(t) \in \big\{0,\ldots, \lceil \sqrt{P} \rceil \big\}$ and $\bar{Y}_{k}^{[m]}(t) \in \mathbb{Z}$, for all $k \in \langle K \rangle$, $m \in \langle M \rangle$ and $t \in \langle n \rangle$, are the inputs and outputs respectively.
The set of channel gains seen by the $k$-th user is given by
$\mathcal{G}_{k} = \bigl\{  G_{ki}^{[m]}(t) : i \in \langle k-1 \rangle, t \in \langle n \rangle, m \in \langle M \rangle  \bigr\}$
and the set of all channel gains is given by $\mathcal{G} = \bigl\{\mathcal{G}_{k} : k \in \langle K \rangle \bigr\}$.
Note that partial CSIT is inherited by the canonical deterministic channel such that
\begin{equation}
\nonumber
G_{ki}^{[m]}(t) = \hat{G}_{ki}^{[m]}(t) + \sqrt{P^{- \alpha_{k}^{[m]}}}\tilde{G}_{ki}^{[m]}(t)
\end{equation}
where the transmitter knows $\hat{G}_{ki}^{[m]}(t)$ but not $\tilde{G}_{ki}^{[m]}(t)$.
For briefness, we use $\bar{X}_{k}^{[m]}$ with a suppressed time index to denote the sequence  $\bar{X}_{k}^{[m]}(1),\ldots, \bar{X}_{k}^{[m]}(n)$, and $\bar{X}_{k}$ with a suppressed subchannel index to denote $\bar{X}_{k}^{[1]},\ldots,\bar{X}_{k}^{[M]}$.
Similarly, we define the received sequences $\bar{Y}_{k}^{[m]}$  and $\bar{Y}_{k}$.
Next, we obtain the different of entropy terms from
\begin{align}
\nonumber
nR_{k} & \leq I\left( W_{k}; \bar{Y}_{k} \mid W_{\langle k+1:K \rangle},\mathcal{G} \right) + o(n) \\
\nonumber
& = H\left( \bar{Y}_{k} \mid W_{\langle k+1:K \rangle},\mathcal{G} \right) -
H\left( \bar{Y}_{k} \mid W_{\langle k:K \rangle},\mathcal{G} \right) + o(n)
\end{align}
where $W_{\langle i:j \rangle} = W_{i},\ldots,W_{j}$. By ignoring the $o(n)$ term and adding all rate bounds, we have
\begin{multline}
\label{eq:n_sum_R_k_upperbound}
n\sum_{k=1}^{K}R_{k}  \leq  \frac{n M}{2} \log \left( P \right) +n o\left( \log \left( P \right) \right) \\
+ \sum_{k=2}^{K}
\underbrace{H\left( \bar{Y}_{k-1} \mid W_{\langle k:K \rangle},\mathcal{G}\right) -
H\left( \bar{Y}_{k} \mid W_{\langle k:K \rangle},\mathcal{G}\right)}_{=H_{k}^{\Delta}}.
\end{multline}
The problem reduces to bounding the differences of entropy terms, $H_{k}^{\Delta}$ for all $k \in \langle K \rangle$, in a DoF sense.
\begin{lemma}
$H_{k}^{\Delta}$ is upperbounded in the DoF sense as
\begin{equation}
\label{eq:delta_k upperbound}
\limsup_{P \rightarrow \infty} \limsup_{n \rightarrow \infty} \frac{H_{k}^{\Delta}}{\frac{nM}{2}\log(P)} \leq \frac{1}{M}\sum_{m=1}^{M} \alpha_{k}^{[m]} = \alpha_{k}.
\end{equation}
\end{lemma}
The proof uses the AIS approach of Davoodi and Jafar \cite{Davoodi2016} and can be found in the Appendix.
By combining the bounds in \eqref{eq:n_sum_R_k_upperbound} and \eqref{eq:delta_k upperbound}, the upperbound in Theorem \ref{theorem:sum DoF} directly follows.
\section{Discussion}
The result in Theorem \ref{theorem:sum DoF} can be extended and generalized in multiple directions. Here we give insights into some of these directions and discuss potential challenges.
\subsection{The Whole DoF Region}
A natural extension to Theorem \ref{theorem:sum DoF} is the characterization of the whole DoF region.
An outerbound for the region is constructed by bounding the sum-DoF for all possible subsets of users
(i.e. $\mathcal{S} \subseteq \langle K \rangle$) in a manner similar to \eqref{eq:sum DoF}.
We denote such outerbound by $\mathcal{D}_{\mathrm{out}}$.
On the other hand, an innerbound can be obtained by separate coding over subchannels.
Considering the $m$-th subchannel as a stand-alone network, the optimum DoF region was
fully characterized in \cite{Piovano2017}, and its achievability is based on rate-splitting with flexible power allocations.
Denoting this region by $\mathcal{D}^{[m]}$,
the innerbound region for the considered parallel channel is described by the Minkowski sum
$\mathcal{D}_{\mathrm{in}} = \frac{1}{M} \left( \mathcal{D}^{[1]} + \cdots + \mathcal{D}^{[M]}\right)$.
The main challenge is to show that this Minkowski sum coincides with $\mathcal{D}_{\mathrm{out}}$ under the total order in Definition \ref{def:total order}.
We observe that $\mathcal{D}_{\mathrm{out}}$ and $\mathcal{D}^{[1]},\ldots,\mathcal{D}^{[m]}$ are all described by sum-DoF inequalities only (no weighted sum-DoF inequalities are necessary).
For the special case where CSIT qualities are uniform in each subchannel, i.e. $\alpha_{i}^{[m]} = \alpha_{k}^{[m]}$ for all $i,j \in \langle K \rangle$ in each $m \in \langle M \rangle$, we have $\mathcal{D}_{\mathrm{out}} = \mathcal{D}_{\mathrm{in}}$.
This follows by noting that each of $\mathcal{D}^{[1]},\ldots,\mathcal{D}^{[m]}$ is a polymatroid, and hence, the linear inequalities that describe their Minkowski sum are obtained by directly summing the corresponding linear inequalities describing the summands \cite[Th. 44.6]{Schrijver2002}\footnote{This result on the Minkowski sum of polymatroids was recently highlighted by Sun and Jafar in \cite{Sun2016} where they referred to \cite[Th. 3]{McDiarmid1975}. We find the statement of the summability result in \cite[Th. 44.6]{Schrijver2002} more accessible.}.
However, when the above uniformity condition is not satisfied, this property cannot be directly applied as the summands are not necessarily polymatroidal.
Showing that $\mathcal{D}_{\mathrm{out}} = \mathcal{D}_{\mathrm{in}}$ under total order requires showing that the summability property of polymatroids in \cite[Th. 44.6]{Schrijver2002} holds for a wider class of bounded polyhedra, which includes $\mathcal{D}^{[1]},\ldots,\mathcal{D}^{[m]}$.
\subsection{Relaxing the Total Order Condition}
A more challenging extension to the result at hand involves the relaxation of the total order condition in Definition \ref{def:total order}.
In this case, the upperbound in Theorem \ref{theorem:sum DoF} is loose in general.
This follows from the observations made in \cite{Rassouli2016}, where it was shown that in the alternating CSIT setting with $K \geq 3$ users, the DoF region is not fully characterized by the marginal probabilities (or the average CSIT qualities in our case) in general.
A conceptually representative example is the one given in \cite[Fig. 3]{Rassouli2016}, in which a $3$-user MISO BC with $3$ subchannels and CSIT qualities $\bm{\alpha}^{[1]} = (1,0,0)$,
$\bm{\alpha}^{[2]} = (0,1,0)$ and $\bm{\alpha}^{[3]} = (0,0,1)$ is considered.
The average qualities are given by $\alpha_{1} = \alpha_{2} = \alpha_{3} = \frac{1}{3}$
from which the sum-DoF upperbound in Theorem \ref{theorem:sum DoF} is equal to $\frac{5}{3}$.
However, a refined upperbound in \cite[Sec. V]{Rassouli2016} yields $\frac{8}{5}$, which is tighter than $\frac{5}{3}$.
It is worth noting that for this example, the maximum sum-DoF achieved through separate coding is $1$.
Hence, any chance of achieving $\frac{8}{5}$ would rely on joint coding across the subchannels.
Another example which shows the strict superiority of joint coding over separate coding in a $3$-user MISO BC with $2$ subchannels where total order does not hold is given in \cite[Fig. 8]{Rassouli2016}.
From the above discussion, it can be concluded that addressing the general non-ordered case requires both new upperbounds and innerbounds.
For the former, it is not clear yet if a straightforward application of the AIS approach \cite{Davoodi2016} (or the newly introduced sum-set inequalities \cite{Davoodi2017}) is sufficient.
As for the latter, whilst a total order does not hold in general, it may be helpful to exploit any partial orders in which total orders hold for subsets of users.
Moreover, using a $\mathrm{PN}$ decomposition similar to the one in Section \ref{subsec:PN decomposition} and applying zero-forcing precoding over known channel vectors, the problem seems to relate to a form of Topological Interference Management (TIM) with alternating connectivity \cite{Sun2013}.
Exploring such parallels is a topic for future research.
\section{Conclusion}
In this paper, we introduced the concept of total order in the parallel MISO BC with arbitrary partial CSIT levels.
We showed that total order yields sum-DoF separability and achieves the maximum possible sum-DoF given average CSIT qualities.
Implications, insights and possible extensions of the result in this paper were thoroughly discussed.
\appendix
Here we give a proof for the bound in \eqref{eq:delta_k upperbound} by following the steps of the AIS approach in \cite{Davoodi2016,Davoodi2016a,Davoodi2016c}.
For notational convenience, we introduce $G_{kk}^{[m]}(t) = 1$ for all $k \in \langle K \rangle$, $t \in \langle n \rangle$ and $m \in \langle M \rangle$.
Due to the  non-degenerate channel model, there exists a constant $\Delta $ such that $0 < \Delta^{-1} \leq |G_{ki}^{[m]}(t)| \leq \Delta  < \infty$ for all $k \in \langle K \rangle$, $i \in \langle k \rangle$, $t \in \langle n \rangle$ and $m \in \langle M \rangle$.
Conditioned on the CSIT, the non-degenerate channel uncertainty implies that the peak of the probability density function behaves as $O\bigl( \bar{P}^{\alpha_{k}^{[m]}} \bigr)$ \cite{Davoodi2016a}, where $\bar{P} = \sqrt{P}$.
\subsubsection{Functional Dependence}
Given the channel $\mathcal{G}_{k-1}$ of user $k-1$, the mapping from the received signal
$\bar{Y}_{k-1}$ to one of the codewords $(\bar{X}_{1},\ldots,\bar{X}_{k})$, which is denoted by $\mathcal{L}$, is random in general.
Similar to \cite{Davoodi2016}, we fix a deterministic mapping
$(\bar{X}_{1},\ldots,\bar{X}_{k}) = \mathcal{L}_{0}\bigl(\bar{Y}_{k-1}, \mathcal{G}_{k-1}\bigr)$ such that
the term $H(\bar{Y}_{k} \mid W_{\langle k:K \rangle},\mathcal{G},\mathcal{L})$ is minimized.
This does not influence the term $H(\bar{Y}_{k-1} \mid W_{\langle k:K \rangle}, \mathcal{G})$, and hence provides an outer bound on $H_{k}^{\Delta}$.
As a result, we proceed while assuming that $\bar{Y}_{k}$ is a function of $\bar{Y}_{k-1}$ and $\mathcal{G}$, i.e.
$\bar{Y}_{k}\bigl(\bar{Y}_{k-1}, \mathcal{G}\bigr)$.
\subsubsection{Aligned Image Sets and Difference of Entropies}
For a given channel realization $\mathcal{G}$, this is defined as the set of all $\bar{Y}_{k-1}$ that have the same image in  $\bar{Y}_{k}$, i.e.
\begin{equation}
\nonumber
\mathcal{A}_{\nu} \left(\mathcal{G} \right) \triangleq
\Bigl\{  \bar{y}_{k-1}  \in \bigl\{\bar{Y}_{k-1} \bigr\} : \bar{Y}_{k}  \left( \bar{y}_{k-1}, \mathcal{G} \right) =
\bar{Y}_{k} \left( \nu, \mathcal{G}  \right)  \Bigr\}.
\end{equation}
Following the same steps in \cite{Davoodi2016}, $H_{k}^{\Delta}$ is bounded in terms of the average size of the aligned image sets as
\begin{equation}
\label{eq:H_delta upperbound}
H_{k}^{\Delta} \leq \log \Bigl( \E \bigl( \big| \mathcal{A}_{\bar{Y}_{k-1}} \left(\mathcal{G} \right) \big| \bigr) \Bigr)
\end{equation}
and the problem becomes to bound this average size.
Note that for a given realization $\nu$ of $\bar{Y}_{k-1}$, we have
\begin{equation}
\label{eq:average_AIS}
\E \bigl( \big| \mathcal{A}_{\nu} \left(\mathcal{G} \right) \big| \bigr) =
\sum_{\lambda \in  \{\bar{Y}_{k-1}\} } \Prob\bigl( \lambda \in \mathcal{A}_{\nu} \left(\mathcal{G} \right) \bigr).
\end{equation}
Next, we bound the probabilities in \eqref{eq:average_AIS}.
\subsubsection{Probability of Image Alignment}
Given $\mathcal{G}_{k-1}$, consider two distinct realizations of $\bar{Y}_{k-1}$ denoted by $\lambda$ and $\nu$, which are produced by the two realizations of $(\bar{X}_{1},\ldots,\bar{X}_{k})$ given by $(\bar{\lambda}_{1},\ldots,\bar{\lambda}_{k}) = \mathcal{L}_{0}\bigl(\lambda, \mathcal{G}_{k-1}\bigr)$
and $(\bar{\nu}_{1},\ldots,\bar{\nu}_{k}) = \mathcal{L}_{0}\bigl(\nu, \mathcal{G}_{k-1}\bigr)$ respectively.
We wish to bound the probability of the event that the images of  $(\bar{\lambda}_{1},\ldots,\bar{\lambda}_{k})$ and
$(\bar{\nu}_{1},\ldots,\bar{\nu}_{k}) $ align at user $k$, i.e. $\lambda \in \mathcal{A}_{\nu} \left(\mathcal{G} \right) $.
For such event, we must have
\begin{multline}
\label{eq:aligned_images}
\sum_{i = 1}^{k} \bigl\lfloor G_{ki}^{[m]}(t) \bar{\lambda}_{i}^{[m]}(t) \bigr\rfloor
=
\sum_{i = 1}^{k} \bigl\lfloor G_{ki}^{[m]}(t) \bar{\nu}_{i}^{[m]}(t) \bigr\rfloor \\
\Rightarrow \left| \sum_{i = 1}^{k} G_{ki}^{[m]}(t) \left( \bar{\lambda}_{i}^{[m]}(t) - \bar{\nu}_{i}^{[m]}(t) \right)  \right| \leq k
\end{multline}
for all $t \in \langle n \rangle$ and  $m \in \langle M \rangle$.
Let us focus on the alignment event in \eqref{eq:aligned_images} for the $t$-th channel use of the $m$-th subchannel, i.e. $\lambda^{[m]}(t) \in \mathcal{A}_{\nu^{[m]}(t)} \left(\mathcal{G} \right)$.
This event is a subset of the event specified by
$\bigl| G_{ki}^{[m]}(t) \bigl( \bar{\lambda}_{i}^{[m]}(t) - \bar{\nu}_{i}^{[m]}(t) \bigr)  \bigr| \leq K$
for any $i \in \langle k-1 \rangle $.
Hence, for $\bar{\lambda}_{i}^{[m]}(t) \neq \bar{\nu}_{i}^{[m]}(t)$, the probability $\Prob\bigl( \lambda^{[m]}(t) \in \mathcal{A}_{\nu^{[m]}(t)} \left(\mathcal{G} \right) \bigr) $ is bounded by the probability of $G_{ki}^{[m]}(t)$ taking values within an interval of length no more than $\frac{2K}{\bar{ | \lambda}_{i}^{[m]}(t) - \bar{\nu}_{i}^{[m]}(t) |}$,
which in turn is bounded by $\frac{2Kf_{\max} \bar{P}^{\alpha_{k}^{[m]}}}{\bar{ | \lambda}_{i}^{[m]}(t) - \bar{\nu}_{i}^{[m]}(t) |}$, where
$f_{\max}$ is the finite positive constant associated with the bounded density assumption.
Since this bound holds for any $i \in \langle k-1 \rangle $, the probability of alignment for the $t$-th channel use of the $m$-th subchannel is no more than
$\frac{2Kf_{\max} \bar{P}^{\alpha_{k}^{[m]}}}{\max_{i \in \langle k-1 \rangle }  | \bar{\lambda}_{i}^{[m]}(t) - \bar{\nu}_{i}^{[m]}(t) |}$.
As in \cite{Davoodi2016,Davoodi2016c}, to bound the probability of alignment in terms of the realizations of $\bar{Y}_{k-1}^{[m]}(t)$, i.e. $\lambda^{[m]}(t)$ and $\nu^{[m]}(t)$, it can be shown that
\begin{equation}
\nonumber
\max_{i \in \langle k-1 \rangle }  | \bar{\lambda}_{i}^{[m]}(t) - \bar{\nu}_{i}^{[m]}(t) | \geq \frac{| \lambda^{[m]}(t) - \nu^{[m]}(t)| - K}
{K\Delta }
\end{equation}
whenever $| \lambda^{[m]}(t) - \nu^{[m]}(t)| > K$.
By letting $L^{[m]}(t)$ denote $| \lambda^{[m]}(t) - \nu^{[m]}(t)| - K$, we have
\begin{equation}
\nonumber
\Prob\bigl( \lambda^{[m]}(t) \in \mathcal{A}_{\nu^{[m]}(t)} \left(\mathcal{G} \right) \bigr) \leq
\begin{cases}
       \frac{2K^{2} \Delta  f_{\max} \bar{P}^{\alpha_{k}^{[m]}}}{ L^{[m]}(t)},
        \ L^{[m]}(t) > 0 \\
       1, \ \text{otherwise}.
\end{cases}
\end{equation}
Moving towards multiple channel uses and subchannels, the probability of alignment is bounded by
\begin{multline}
\label{eq:probability_image_alignment}
\Prob\bigl( \lambda \in \mathcal{A}_{\nu} \left(\mathcal{G} \right) \bigr)
\leq  \Bigl( \max\bigl\{ 2K^{2} \Delta  f_{\max},1 \bigr\}  \Bigr)^{nM}\\
  \times \prod_{m =1}^{M}  \bar{P}^{n\alpha_{k}^{[m]}}
\times \prod_{m = 1}^{M}  \prod_{t:L^{[m]}(t) > 0} \frac{1}{ L^{[m]}(t) }.
\end{multline}
\subsubsection{Bounding the Average Size of Aligned Image Sets and Combining Bounds}
From \eqref{eq:average_AIS} and \eqref{eq:probability_image_alignment}, we have
\begin{align}
\nonumber
\E \bigl( \big| \mathcal{A}_{\nu} \left(\mathcal{G} \right) \big| \bigr)
&\leq  \Bigl( \max\bigl\{ 2K^{2} \Delta  f_{\max},1 \bigr\}  \Bigr)^{nM}  \bar{P}^{n\sum_{m=1}^{M}\alpha_{k}^{[m]}} \\
\nonumber
\times \prod_{m=1}^{M} \prod_{t=1}^{n}\Biggl(
& \sum_{\lambda^{[m]}(t):L^{[m]}(t)  \leq  0} 1 +
\! \! \! \! \! \!
\sum_{\lambda^{[m]}(t): 0 < L^{[m]}(t)  \leq  Q_{y}} \! \! \frac{1}{ L^{[m]}(t) } \Biggr) \\
\nonumber
& \leq \Bigl( \max\bigl\{ 2K^{2} \Delta  f_{\max},1 \bigr\}  \Bigr)^{nM} \bar{P}^{n\sum_{m=1}^{M}\alpha_{k}^{[m]}}   \\
\label{eq:average_AIS_bound}
& \quad \quad \quad \quad  \quad
\times \Bigl( \log(\bar{P}) + o\bigl( \log(\bar{P}) \bigr) \Bigr)^{nM}
\end{align}
where $Q_{y} = \lceil \bar{P} \rceil  K \Delta  $.
The bound in \eqref{eq:average_AIS_bound} is obtained using the techniques in  \cite{Davoodi2016,Davoodi2016c}.
This bound holds for all $\nu \in \mathcal{A}_{\nu}$.
Combining this with \eqref{eq:H_delta upperbound}, we obtain \eqref{eq:delta_k upperbound}.
\begin{remark}
The main difference compared to \cite{Davoodi2016} is encompassed in \eqref{eq:probability_image_alignment} and
\eqref{eq:average_AIS_bound}.
Due to the parallel subchannels, the CSIT quality changes throughout the transmission of one codeword. This is exhibited in the term
$\bar{P}^{n\sum_{m=1}^{M}\alpha_{k}^{[m]}}$, which  translates into the average CSIT quality in \eqref{eq:delta_k upperbound}.
\end{remark}
\section*{Acknowledgment}
The authors gratefully acknowledge insightfull discussions with Arash Gholami Davoodi (University of California, Irvine) regarding the aligned image sets approach.
\bibliographystyle{IEEEtran}
\bibliography{References}

\end{document}